\documentclass[useAMS,usenatbib]{mn2e}

\usepackage{graphicx}
\usepackage{txfonts}
\usepackage[authoryear]{natbib}
\usepackage{subfigure}

\title[Absolute properties of the main-sequence eclipsing binary FM Leo]{Absolute properties of the main-sequence eclipsing binary FM Leo}
\author[M. Ratajczak et al.]
{M. Ratajczak,$^{1}$\thanks{E-mail:milena@ncac.torun.pl}
 T. Kwiatkowski,$^{2}$ A. Schwarzenberg-Czerny,$^{2,3}$ W. Dimitrov,$^{2}$ \newauthor
M. Konacki,$^{1,2}$ K.G. He{\l}miniak,$^{1}$ P. Bartczak,$^{2}$ M. Fagas,$^{2}$ K. Kami\'nski,$^{2}$ \newauthor 
P. Kankiewicz,$^{4}$ W. Borczyk$^{2}$ and A. Ro{\.z}ek$^{2}$\\
$^{1}$Department of Astrophysics, Nicolaus Copernicus Astronomical Center, Rabia\'nska 8, 87-100 Toru\'n, Poland\\
$^{2}$Astronomical Observatory, Adam Mickiewicz University, S$\rm\l$oneczna 36, 60-186 Pozna\'n, Poland\\
$^{3}$Nicolaus Copernicus Astronomical Center, Bartycka 18, 00-716 Warsaw, Poland\\
$^{4}$Astrophysics Division, Institute of Physics, Jan Kochanowski University, \'Swi\c etokrzyska 15, 25-406 Kielce, Poland}

\begin{document}

\date{Accepted ... Received ...; in original form ...}

\pagerange{\pageref{firstpage}--\pageref{lastpage}} \pubyear{2009}

\maketitle

\label{firstpage}

\begin{abstract}
First spectroscopic and new photometric observations of the eclipsing binary FM~Leo are presented. The main aims were to determine orbital and stellar parameters of two components and their evolutionary stage. First spectroscopic observations of the system were obtained with DDO and PST spectrographs. The results of the orbital solution from radial velocity curves are combined with those derived from the light-curve analysis (\textit{V}-band ASAS-3 photometry and supplementary observations of eclipses with 1~m and 0.35~m telescopes) to derive orbital and stellar parameters. \textit{JKTEBOP}, Wilson-Devinney binary modelling codes and a two-dimensional cross-correlation (TODCOR) method were applied for the analysis. We find the masses to be $M_{1}$ = 1.318 $\pm$ 0.007 and $M_{2}$ = 1.287 $\pm$ 0.007 $\rmn M_{\sun}$, the radii to be $R_{1}$ = 1.648  $\pm$ 0.043 and $R_{2}$ = 1.511  $\pm$ 0.049 $\rmn R_{\sun}$ for primary and secondary stars, respectively. The evolutionary stage of the system is briefly discussed by comparing physical parameters with current stellar evolution models. We find the components are located at the main sequence, with an age of about 3~Gyr.
\end{abstract}


\section{Introduction}
\label{sec1}

Binary stars are the main source of fundamental data on stellar masses and radii. High quality and quantity of binaries' observations can help in testing of stellar evolution theory. Accurate masses and radii (determined better than 3~$\%$) derived from detached binary systems' analysis have led to a number of new and interesting results on the properties and evolution of normal stars \citep{tor}. Such data can be used to improve the treatment of convection, diffusion, and other non-classical effects which are essential to our understanding of stellar structure.

	FM~Leo, also known as a HD~97422, HIC~54766 is an EA-type eclipsing binary. Its extensive photometry was listed by ASAS \citep{pojm} under the code ASAS~111245+0020.9. Its maximum magnitude comes to $V_{\rm max} = 8.47$~mag. An analysis of the light-curve reveals two eclipses of nearly identical depth and a flat maximum proving that FM~Leo is a detached binary consisting of two nearly identical spherical components.

   An analysis of ASAS-3 \citep{pojm} and Hipparcos \citep{pery} databases yields an orbital period of $6 \fd 72863$ \citep{otero}. The star is classified as F8 \citep{pery}. The Hipparcos parallax of the binary is $\pi = 8.35 {\pm}$ 1.17~mas.

   In this paper we present results of the exact determination of stellar parameters of FM Leo. Since there was no radial velocity (RV) curve for this star in literature, we resorted to our own observations. We also present new photometric observations of eclipses of the binary. Our data and reduction procedures are described in Sect.~\ref{sec2}. In Sect.~\ref{sec3} we analyse observations using \textit{JKTEBOP} (photometry) \citep{south04a,south04b} and our own procedure (spectroscopy) to find the solution for the system and present absolute properties of FM~Leo, while Sect.~\ref{sec4} is devoted to discussion on the evolutionary status and the age of the system.

\section{Data and reduction}
\label{sec2}

\subsection{Spectroscopy}

Spectroscopic observations were performed with the Cassegrain-focus spectrograph at the 1.9~m telescope of the David Dunlap Observatory (DDO) in 28~April -- 4~May, 2006. Typical exposure time was 20~min. The spectra in number of 12 cover the wavelength range of 5100 -- 5275 $\AA$ at a resolving power of $R \approx 20\,000$. They were obtained with signal-to-noise $\rm S/N$ ratios ranging from 30 to 40.

Additional 16 higher resolution ($R \approx 35000$) spectra were obtained during the period 31~March -- 25~May,~2008 with the Pozna\'n Spectroscopic Telescope (PST) -- the binary telescope using one of 0.4 m mirrors of~a Newtonian focus, connected by optic fibers with an echelle spectrograph \citep{baranowski}. Typical exposure time was 30~min. The spectra cover the full range of 4480 -- 9250 $\AA$ and consist of 64 orders. Typical signal-to-noise ratio is $\rm S/N$~$\approx$~30. The observations were performed during bad weather conditions. All radial velocity (RV) measurements are presented in Table~\ref{tab_rv}.

The data reduction was performed with IRAF package\footnote{\textit{IRAF} is written and supported by the \textit{IRAF} programming group at the National Optical Astronomy Observatories (NOAO) in Tucson, AZ. NOAO is operated by the Association of Universities for Research in Astronomy (AURA), Inc. under cooperative agreement with the National Science Foundation. \tt{http://iraf.noao.edu/}} following the standard procedures, i.e. background subtraction, flatfielding, wavelength calibration using the emission lines of a Th-Ar lamp, and normalisation to the continuum through a polynomial fit. 

In order to determine radial velocities of components, one- and two-dimensional cross-correlation techniques have been used. We derived RV curves for every component using both methods.

To determine radial velocity curves of the system the spectra were cross-correlated against various template spectra and by using different procedures. In the case of one-dimensional correlation we took as a template the spectrum of HD~102870, which is of similar spectral type (F9V) as FM~Leo. The template spectrum was obtained during the same observing run as for FM~Leo. The radial velocities were measured using FXCOR task in IRAF. In the case of two-dimensional correlation we used synthetic spectra created by the ATLAS9 \citep{kurucz} for stars with effective temperatures of $\rm T_1 = 6300$~K and $\rm T_2 = 6200$~K,  solar metallicity, and rotational velocities of $\rm v_1 = 10$ km~s$^{-1}$ and $\rm v_2 = 12$ km~s$^{-1}$, respectively. The cross-correlation was performed with the TODCOR method \citep{zucker}.

Mean formal error of measurements in the case of two-dimensional correlation $\overline \sigma = 0.7$~km~s$^{-1}$ ($\overline \sigma_{\rm PST}~=~0.4~$~km~s$^{-1}$~and~$\overline \sigma_{\rm DDO} = 1.1$~km~s$^{-1}$) was smaller than in the case of one-dimensional correlation ($\overline \sigma=1.4$~km~s$^{-1}$), so we took the TODCOR result as the final one.

\begin{figure}
  \includegraphics[angle=270,width=\columnwidth]{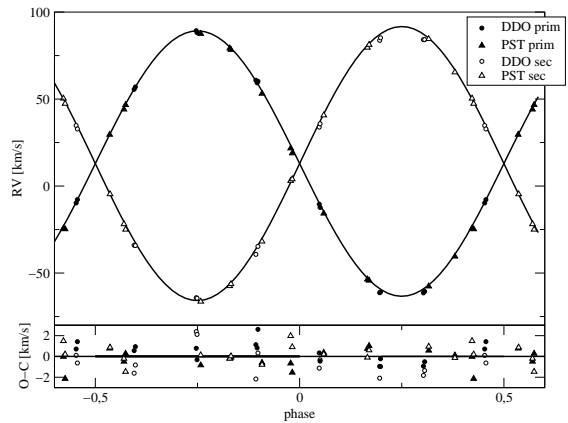}
  \caption{Radial velocity curves of FM~Leo derived by using the two-dimensional correlation method. The solid lines in the top panel are the best-fitting synthetic curves. The bottom panel shows the residuals of the fits.}    \label{fm_rv}
\end{figure}

\subsection{Photometry}

FM~Leo was first identified as an EA-type eclipsing binary with a maximum, out-of-eclipse \textit{V}-band magnitude of $V_{\rm max}~=~8.467~\pm~0.016$~mag in the Hipparcos catalogue \citep{pery}.

A total of 324 \textit{V}-band measurements of the object were gathered between November 2000 and July 2004 during the ASAS-3 survey \citep{pojm}. Mean error of measurements was $\overline{\sigma} = 0.017$~mag. The resulting light-curve exhibits periodic eclipses with depth of $0.5$~mag.

We supplemented the long-term photometry from ASAS with additional measurements of moments of eclipses. 46 \textit{V}-filter observations with exposure time of 40~s were obtained with 1~m Elizabeth telescope in SAAO during one run in December,~2008. Supplementary 2655 measurements with exposure time of 5~s obtained with 0.35~m telescope (Kielce) were gathered over 4 observing runs between 14~April and 28~May,~2008. The Schmidt-Cassegrain telescope located in Kielce Observatory is equipped by the SBIG ST-7XE CCD camera, working typically 25 K below the ambient temperature. Focal reducer 0.5x allows to obtain effective focal length of 1800 mm, providing a 13.0~x~8.9~arcmin field of view.
Data were processed with standard data reduction procedures including bias, dark frames subtraction, and flat-fielding. Typically we used median-combined frames composed from 5 or 7 bias/dark/flat images.
The aperture photometry procedure provided by PHOTOM - STARLINK package\footnote{\tt{http://starlink.jach.hawaii.edu/starlink}} was used to measure instrumental magnitudes. Next, using one comparison star (HD 97441) we estimated the differential
magnitude and used values averaged over every 5 measurements for further analysis. Mean error of Kielce observations was $\overline \sigma=0.018$~mag, while for SAAO measurements we obtained $\overline \sigma= 0.017$~mag.

\section{Analysis}
\label{sec3}

The preliminary orbit solution and physical parameters of FM~Leo have been estimated using \textit{PHOEBE} (PHysics Of Eclipsing BinariEs) \citep{prsa} which is based on the Wilson-Devinney method. We used that estimation as input data to spectroscopic and photometric analysis tools mentioned in this section.

\subsection{Spectroscopic analysis}

Using the procedure that fits a double-keplerian orbit to radial velocity measurements and minimises the $\chi^2$ function with Levenberg-Marquardt algorithm we obtained RV curves with RMS of fitting for Borowiec data $\sigma_{\rm PST 1}~=~0.88$~km~s$^{-1}$ for the primary and $\sigma_{\rm PST 2}~=~0.97$~km~s$^{-1}$ for the secondary (without 4 measurements close to 0 phase the values are: $\sigma_{\rm PST 1}~=~0.63$~km~s$^{-1}$, $\sigma_{\rm PST 2}~=~0.71$~km~s$^{-1}$), and $\sigma_{\rm DDO 1}~=~0.86$~km~s$^{-1}$, $\sigma_{\rm DDO 2}~=~1$~km~s$^{-1}$ for DDO measurements. As input data (period $P$, semiamplitudes $K_1$, $K_2$ and systemic velocity $\gamma$) we used values obtained from \textit{PHOEBE}. The fit where $e$ was set as a free parameter resulted in eccentricity value undistinguishable from zero (within error limits) and its RMS did not differ significantly from the one when a strictly circular orbit was assumed. Thus for further analysis we kept $e=0$ fixed. We assumed period $P$ and inclination $i$ with errors from the preliminary solution of the light-curve analysis (Sec.~\ref{subsec1}).
 
That orbit solution yields masses of components $M_{1}$, $M_{2}$, semimajor axis $a$, velocity semiamplitudes of $K_{1}$, $K_{2}$, and the system radial velocity $\gamma$ presented in Table~\ref{tab1}. Final radial velocity curves are shown in Fig.~\ref{fm_rv}.

\subsection{Photometric analysis}
\label{subsec1}

 \begin{figure}
   \centering
   \includegraphics[angle=270,width=\columnwidth]{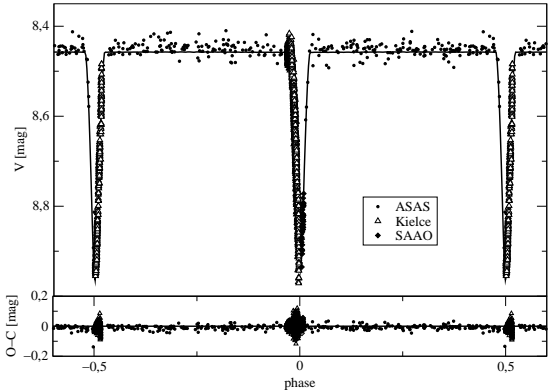}
   \caption{\textit{V} light-curve of FM~Leo. The solid line in the top panel is the best-fitting synthetic light-curve generated by \textit{JKTEBOP}. The bottom panel shows the residuals of the fit.}
              \label{fm_lc}
    \end{figure}

Using \textit{JKTEBOP} procedure by John Southworth\footnote{\tt{http://www.astro.keele.ac.uk/~jkt/codes/jktebop.html}} \citep{south04a,south04b} based on \textit{EBOP} code \citep{popper,etzel} we fit the model to photometric observations. We held the mass ratio $q$ derived from preliminary solution from \textit{PHOEBE} fixed and orbital period $P$, ephemeris timebase $T_0$, light scale factor, orbit inclination $i$, surface brightness ratio, sum and ratio of the radii adjusted with initial values from \textit{PHOEBE}.

Finally we derived the radii of the stars $R_1$, $R_2$, and improved values of period $P$ and ephemeris timebase $T_0$. Results are presented in Table~\ref{tab1}. Additionally we used the same procedure for a bootstrapping error analysis. The final light-curve of FM~Leo is shown in Fig.~\ref{fm_lc}. We obtained RMS of fitting of $\sigma~=~0.02$~mag. 

\subsection{Absolute dimensions}
After using the procedure fitting the orbit to radial velocity measurements and \textit{JKTEBOP} for photometric observations we checked both solutions together in \textit{PHOEBE}. To calculate the absolute dimensions of the eclipsing binary system we used the \textit{JKTEBOP} procedure - \textit{JKTABSDIM}, the code in which careful attention is paid to correctly propagating errors. We took velocity semiamplitudes, orbital eccentricity, periastron longitude, period, orbital inclination, fractional stellar radii, apparent magnitudes, effective temperatures of the stars with uncertainties as input quantities. As a result we derived the absolute masses and radii of two stars, their luminosities and absolute bolometric magnitudes, their surface gravities and synchronous rotational velocities, each with error budget. A full compilation of stellar and orbital parameters of FM~Leo obtained by using mentioned code is presented in Table~\ref{tab1}.

\subsection{Temperatures}
	
To determine individual effective temperatures of both components we used color indices for FM~Leo \citep{cutri} and theoretical color-temperatures relations \citep{bessel}. As a result we obtained the temperature of the primary $T_{\rm eff1}=6316~\pm~240$~K. The secondary's temperature was estimated by using \textit{PHOEBE} with the primary's temperature held fixed. We obtained value of $T_{\rm eff2}=6190~\pm~211$~K for the secondary. The combined photometry used to calculate the temperature of primary leaded to underestimation of its value (combination with the light of a cooler star) but by less than the untertainty.

\begin{table}
 \caption{Absolute dimensions of FM~Leo. Surface gravity acceleration is denoted with $\log{g}$, where $g$ is given in cgs units.}
  \centering
   \begin{tabular}{c c c c}
 \hline
Parameter & Primary & Secondary & Unit \\[0.5ex]
\hline
 \hline
$T_{0}$ & \multicolumn{2}{c}{HJD 2452499.182 $\pm$ 0.002} &  \\ 
$P$ & \multicolumn{2}{c}{6.728606 $\pm$ 0.000006} & d \\
$e$ & \multicolumn{2}{c}{0} &  \\
$\gamma$ & \multicolumn{2}{c}{11.87 $\pm$ 0.13} & km~s$^{-1}$ \\
$i$ & \multicolumn{2}{c}{87.98 $\pm$ 0.06}  & deg \\
$a$ & \multicolumn{2}{c}{20.631 $\pm$ 0.052} & $\rm R_{\sun}$ \\
$K$ & 76.619 $\pm$ 0.273 & 78.463 $\pm$ 0.284 & km~s$^{-1}$ \\
$M$ & 1.318 $\pm$ 0.007 & 1.287 $\pm$ 0.007 & $\rm M_{\sun}$ \\
$R$ & 1.648 $\pm$ 0.043 & 1.511 $\pm$ 0.049 & $\rm R_{\sun}$ \\
$L$ &  1.806 $\pm$ 0.086  &  1.617 $\pm$ 0.102  & $\rm L_{\sun}$ \\
$T_{\rm eff}$ & 6316 $\pm$ 240 & 6190 $\pm$ 211 & K \\
$\log{g}$ & 4.124 $\pm$ 0.023 & 4.189 $\pm$ 0.028 & \\[1ex]

\hline
\end{tabular}
\label{tab1}
 \end{table}

\section{Comparison with stellar models}
\label{sec4}

\begin{figure}
   \centering
   \includegraphics[angle=270,width=\columnwidth]{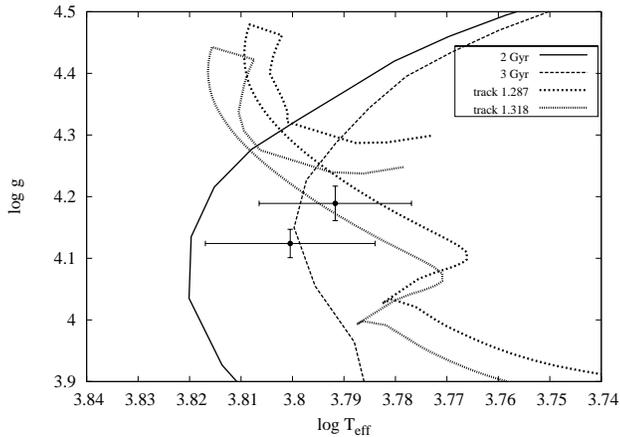}
   \caption{Evolutionary tracks for the measured masses of $1.318 \rm M_{\sun}$ and $1.287 \rm M_{\sun}$ and isochrones of ages 2~Gyr and 3~Gyr generated by Yonsei-Yale code for solar metallicity ($Z = 0.018$) and $\alpha$-enhancement of zero.}
              \label{fig:logT_logg_track}
 \end{figure}

To prove the solution for both components we compared them with current stellar evolution models -- evolutionary tracks for certain masses and checked if a single isochrone fits both stars simultaneously. We interpolated evolutionary tracks at the measured masses of primary ($M_1=1.318 \rm M_{\sun}$) and secondary ($M_2=1.287 \rm M_{\sun}$) component using Yonsei-Yale code \citep{yi} for various metallicities and $\alpha$-enhancement of zero. The best fitting we get for assumed solar metallicity ($Z = 0.018$). The tracks on $\log{T}$ -- $\log{g}$ plane are illustrated in~Fig.~\ref{fig:logT_logg_track}. The parameters of both components are consistent with evolutionary tracks for given masses.

To estimate the age of the system and prove obtained parameters we considered three different sets of theoretical calculations: Yonsei-Yale \citep{yi}, Padova \citep{padova}, and Geneva \citep{geneva} for various ages and metallicities. We obtained the best fitting for ages in the range of 2 -- 4~Gyr and assumed metallicity of $Z = 0.018$.

\begin{figure*}
\centering
\subfigure[]{
\includegraphics[angle=270,width=\columnwidth]{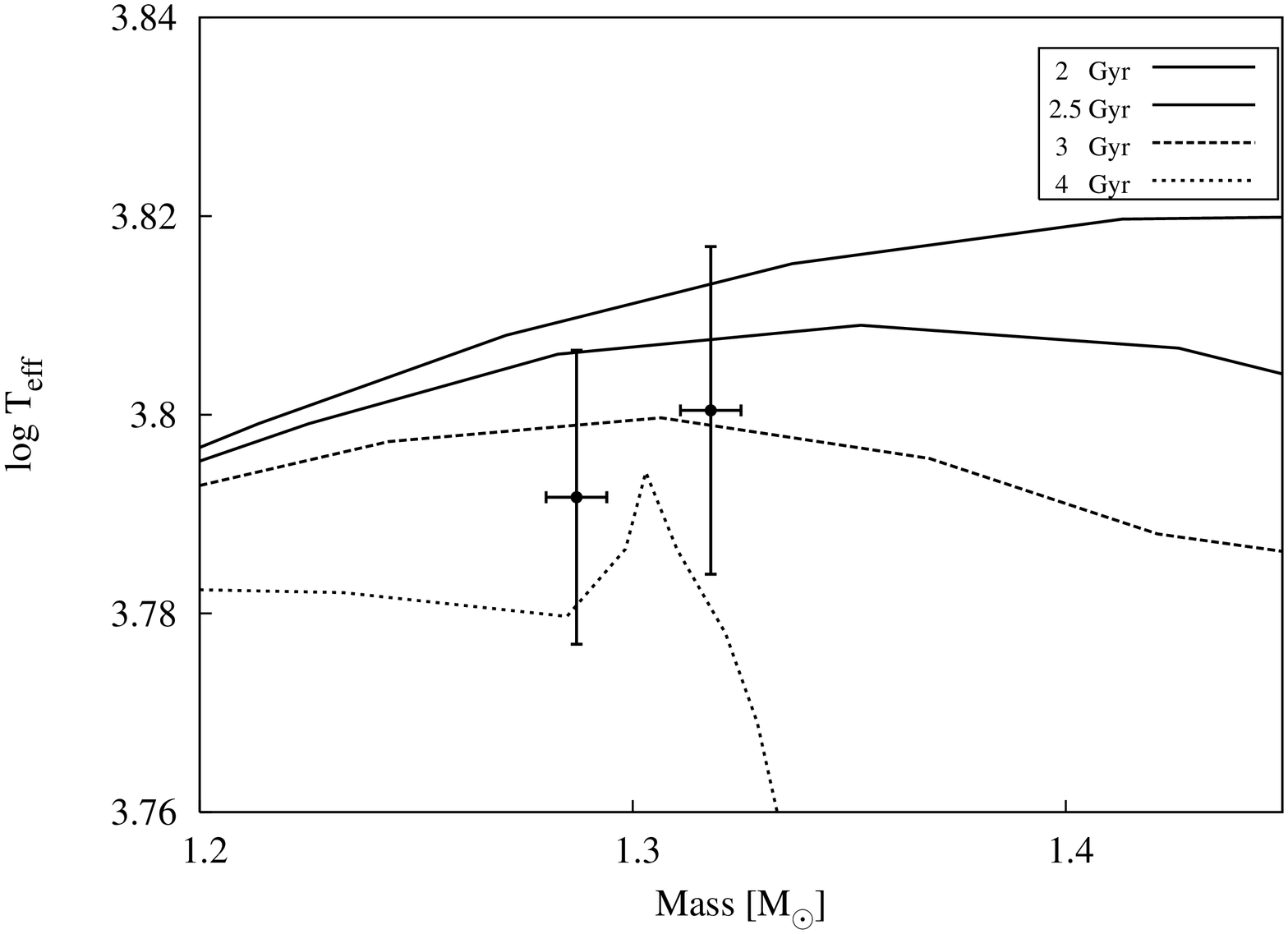}
\label{fig:subfig1}
}
\subfigure[]{
\includegraphics[angle=270,width=\columnwidth]{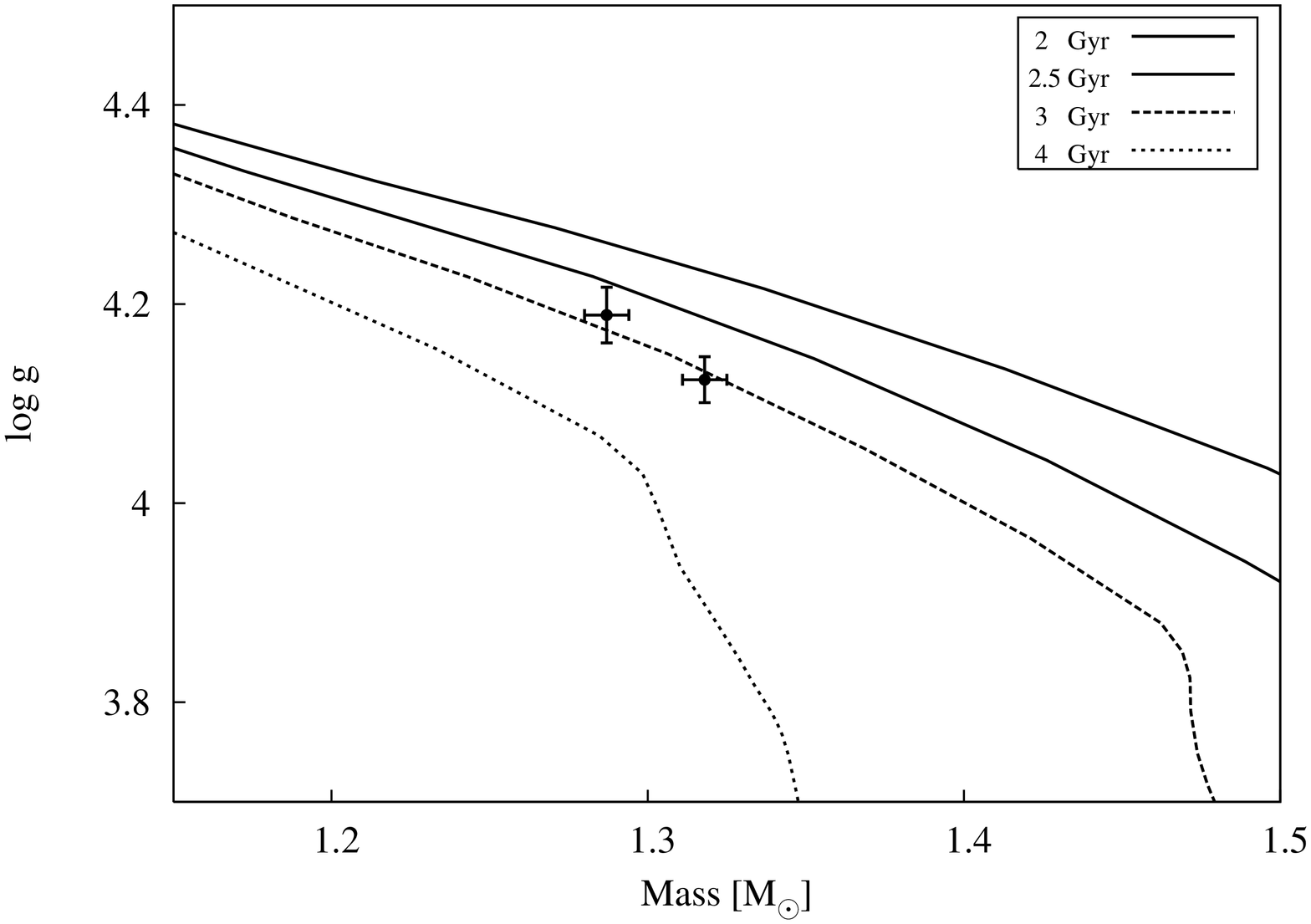}
\label{fig:subfig2}
}
\caption{The location of components of FM~Leo on several planes (\subref{fig:subfig1}: mass -- $\log{T}$ and \subref{fig:subfig2}: mass -- $\log{g}$) is compared with isochrones from Yale-Yonsei \citep{yi} theoretical models. The age for isochrones is 2, 2.5, 3, 4~Gyr and the adopted metallicity is $Z = 0.018$.}
\label{fig:izo}
\end{figure*}

The location of two components of FM~Leo on the various planes (Fig.~\ref{fig:izo}) suggests the age of the system of about 3~Gyr. We used Yonsei-Yale \citep{yi} evolutionary models with improved opacities and equations of state to generate isochrones. Helium diffusion and convective core overshooting have also been taken into consideration during calculations. We assumed metallicity of $Z = 0.018$ and $\alpha$-enhancement of zero.

The location of the components on the mass -- $\log{T}$ (Fig.~\ref{fig:subfig1}) plane suggests the age of the system in the range of 2.5 -- 4~Gyr, the wide extent is caused by considerable temperature error. According to the location of the stellar parameters on the mass -- $\log{g}$ diagram (Fig.~\ref{fig:subfig2}), the 3~Gyr isochrone fits to both components perfectly. 

\section{Conclusions}
\label{sec5}

First spectroscopic and new photometric observations of the eclipsing binary FM~Leo combined with data from literature (ASAS photometry) have allowed us to derive definitive orbital parameters and physical properties of the component stars. The orbital and physical analysis is presented for the first time. With the mass and radius of each component of FM~Leo determined better than 1~$\%$ and 4~$\%$ respectively, we compared the observations with current stellar evolution models interpolated for this system. Using several methods with different input parameters and plotting them on various planes, we find the components are located at the main sequence, with an age of about 3~Gyr. The metallicity of the system is comparable with the solar one. 

To confirm the results we calculated the distance to the system from determined radius, temperature and apparent magnitude and compared it with the value from Hipparcos ($\pi~=~8.35~\pm~1.17~$~mas, $d~=~119.8~\pm~19.5~$~pc) \citep{pery}. We estimated the parallax to the system of $\pi~=~8.55~\pm~2.05$~mas and the distance $d~=~130.4~\pm~25.5$~pc. The estimation of reddening was performed by using maps of dust infrared emission by \citet{schlegel}. We received the color excess of $E(B-V)~=~0.043$. The test confirmed our estimations are acceptable.

To make radii estimations more precise new photometric data of times of eclipses are needed. It would open a potential application of fully characterized FM~Leo for tests of evolutionary tracks and isochrones.

\section*{Acknowledgments}

ASC acknowledges support from the Polish MNiSW grant No.~N~N203~3020~35. MK is supported by
the Foundation for Polish Science through a FOCUS grant and fellowship, by the Polish Ministry of Science
and Higher Education through grant No.~N203~005~32/0449. We are grateful to Heide de Bond and Jim
Thomson for their help in the observations at DDO.

\appendix
\section{Radial velocity measurements}
\label{tab_rv}

\begin{table*}
\caption{Single RV measurements for FM Leo.} 
\label{tab_rv}
\begin{tabular}{ccccc}
\hline

HJD & $\rm v_1$ & $\sigma_{\rm v_1}$ & $\rm v_2$ & $\sigma_{\rm v_2}$ \\
\\
(2450000. + )& [km~s$^{-1}$] & [km~s$^{-1}$] & [km~s$^{-1}$] & [km~s$^{-1}$] \\
\hline
\hline
\multicolumn{5}{c}{}\\
		
3853.6735&    -61.4142&	1.09   & 84.1333	 & 1.03 \\
3853.6902&    -60.5887&	1.14   & 84.1846	 & 1.22 \\
3854.6804&    -9.7157 &	1.04   & 34.7761	 & 0.95 \\
3854.6978&    -7.8169 &	1.06   & 32.8196	 & 1.21 \\
3855.6381&    55.6258 &	1.05   & -34.00368 	 & 1.17 \\
3855.6546&    56.9801 &	1.11   & -34.21597 	 & 1.21 \\
3856.6581&    89.2604 &	0.96   & -64.192	 & 1.32 \\
3856.6740&    88.1706 &	0.95   & -64.47747 	 & 1.43 \\
3857.6752&    60.2235 &	1.03   & -34.6808	 & 0.89 \\
3858.6816&    -10.4293&	0.88   & 33.8878	 & 1.35 \\
3858.6975&    -12.298 &	0.96   & 35.8051	 & 0.97 \\
3859.6785&    -61.3133&	1.11   & 83.7468	 & 1.12 \\
3859.6940&    -60.947 &	0.98   & 85.2303	 & 1.08 \\
4606.354447& -53.631  & 0.117  & 79.604          & 0.169 \\
4606.380649& -54.286  & 0.268  & 81.256          & 0.342 \\
4612.364547& -15.733  & 0.258  & 40.693          & 0.434 \\
4602.344046& 44.082   & 0.392  & -21.850         & 0.315 \\
4602.372331& 46.661   & 0.442  & -24.969         & 0.686 \\
4601.345582& -24.307  & 0.329  & 50.360          & 1.050 \\
4601.371864& -24.761  & 0.262  & 47.36           & 0.559 \\ 
4598.359986& 21.701   & 0.417  & 3.060           & 0.244 \\
4598.387449& 18.873   & 0.371  & 3.986           & 0.240 \\
4597.353753& 79.155   & 0.352  & -57.431         & 0.273 \\
4597.379573& 78.350   & 0.300  & -56.290         & 0.390 \\ 
4595.381744& 29.562   & 0.557  & -4.624          & 0.330 \\
4594.336033& -40.459  & 0.698  & 65.418          & 0.314 \\
4584.430998& 53.129   & 0.243  & -31.815         & 0.238 \\
4583.419625& 87.563   & 0.256  & -66.370         & 0.172 \\
4580.449742& -57.627  & 0.154  & 84.575          & 0.283 \\
4557.414037& 60.804   & 0.175  & -39.276         & 0.215 \\
\hline
\end{tabular}
\end{table*}

\bsp

\label{lastpage}


\begin{thebibliography}{}

\bibitem[\protect\citeauthoryear{Baranowski et al.}{2009}]{baranowski}
Baranowski, R., et al., 2009, 
MNRAS, 396, 2194 

\bibitem[\protect\citeauthoryear{Bessel, Castelli \& Plez}{1998}]{bessel}
Bessel, M.~S., Castelli, F., \& Plez, B., 1998,
A\&A, 333, 231

\bibitem[\protect\citeauthoryear{Castelli \& Kurucz}{2003}]{kurucz}
Castelli, F., \&  Kurucz, R.~L., 2003,
Modelling of Stellar Atmospheres, 210, 20P

\bibitem[\protect\citeauthoryear{Cutri et al.}{2003}]{cutri}
Cutri, R.~M., et al., 2003,
The IRSA 2MASS All-Sky Catalog of Point Sources, NASA/IPAC Infrared Science Archive. ~http://irsa.ipac.caltech.edu/applications/Gator/

\bibitem[\protect\citeauthoryear{Etzel}{1981}]{etzel}
Etzel, P.~B., 1981,
Photometric and Spectroscopic Binary Systems, 11


\bibitem[\protect\citeauthoryear{Lejeune \& Schaerer}{2001}]{geneva}
Lejeune, T., \& Schaerer, D., 2001,
A\&A, 366, 538

\bibitem[\protect\citeauthoryear{Marigo et al.}{2008}]{padova}
Marigo, P., Girardi, L., Bressan, A., Groenewegen, M.~A.~T, Silva, L., \& Granato, G.~L., 2008,
A\&A, 482, 883

\bibitem[\protect\citeauthoryear{Otero}{2003}]{otero}
Otero, S.~A., 2003,	
Information Bulletin on Variable Stars, 5480, 1

\bibitem[\protect\citeauthoryear{Perryman et al.}{1997}]{pery}
Perryman, M.~A.~C. et al., 1997,
A\&A, 323, L49

\bibitem[\protect\citeauthoryear{Pojma\'{n}ski}{2002}]{pojm}
Pojma\'{n}ski, G., 2002,
Acta Astron., 52, 397


\bibitem[\protect\citeauthoryear{Popper \& Etzel}{1981}]{popper}
Popper, D.~M., \& Etzel, P.~B., 1981
AJ, 86, 102

\bibitem[\protect\citeauthoryear{Pr{\v s}a \& Zwitter}{2005}]{prsa}
Pr{\v s}a, A., \& Zwitter, T.\ 2005,
ApJ, 628, 426


\bibitem[\protect\citeauthoryear{Schlegel, Finkbeiner \& Davis}{1998}]{schlegel}
Schlegel, D.~J., Finkbeiner, D.~P., \& Davis, M., 1998,
ApJ, 500, 525

\bibitem[\protect\citeauthoryear{Southworth, Maxted \& Smalley}{2004a}]{south04a}
Southworth, J., Maxted, P.~F.~L, \& Smalley, B., 2004a,
MNRAS, 351, 1277

\bibitem[\protect\citeauthoryear{Southworth et al.}{2004b}]{south04b}
Southworth, J., Zucker, S., Maxted, P.~F.~L, \& Smalley, B., 2004b,
MNRAS, 355, 986

\bibitem[\protect\citeauthoryear{Torres, Andersen \& Gim{\'e}nez}{2009}]{tor}
Torres, G., Andersen, J., Gim{\'e}nez, A., 1994,
preprint, astro-ph/arXiv:0908.2624

\bibitem[\protect\citeauthoryear{Yi et al.}{2004b}]{yi}
Yi, S., Demarque, P., Kim, Y.~C., Lee, Y.~W., Ree, C.~H., Lejeune, T., \& Barnes, S., 2001,
ApJS, 136, 417

\bibitem[\protect\citeauthoryear{Zucker \& Mazeh}{1994}]{zucker}
Zucker, S., Mazeh, T., 1994,
ApJ, 420, 806

\end{thebibliography}
\end{document}